\documentclass[fleqn,twoside]{article}
\usepackage{espcrc2}
\usepackage{graphicx}
\usepackage[figuresright]{rotating}

\newcommand{\AmS}{{\protect\the\textfont2
  A\kern-.1667em\lower.5ex\hbox{M}\kern-.125emS}}

\title{Proton spin structure and the axial U(1) problem}
\author{Steven D. Bass
\addressmark\thanks{
 Invited talk at the Workshop on the
 Spin Structure of the Proton
 and Polarized Collider Physics, ECT* Trento, July 23-28, 2001}
 \address[ECT]{ECT*, Strada delle Tabarelle 286, I-38050
        Villazzano, Trento, Italy}}

\begin{document}

\begin{abstract}
We emphasise the relation between the spin structure of the proton and
the axial U(1) problem.
New experiments motivated by the proton spin problem which could shed
light on the nature of U(1) symmetry breaking in QCD are discussed.
\vspace{1pc}
\end{abstract}

\maketitle

\section{INTRODUCTION}

The spin structure of the proton and $\eta'$ physics provide complementary 
windows on the role of gluonic degrees of freedom in dynamical $U_A(1)$ 
symmetry breaking in QCD.
The small value of the flavour-singlet axial charge $g_A^{(0)}$ which is 
extracted from the first moment of $g_1$ 
(the nucleon's first spin dependent structure function) \cite{windmolders}
\begin{equation}
\left. g^{(0)}_A \right|_{\rm pDIS} = 0.2 - 0.35
\end{equation}
and the large mass of the $\eta'$ meson point to large violations of OZI 
in the flavour-singlet $J^P=1^+$ channel
\cite{bass99,cracow,okubo}.
The strong (QCD) axial anomaly is central to theoretical explanations of 
the small value of $g_A^{(0)}|_{\rm pDIS}$
(about 50\% of the OZI value 0.6) and the large $\eta'$ mass.
In this paper we review the role of the anomaly in the spin structure of 
the proton and highlight the interplay between the proton's internal spin 
structure and axial U(1) dynamics.

In the QCD parton model the proton's flavour-singlet axial charge 
$g_A^{(0)}$ receives contributions from the amount of spin carried 
by quark and gluon partons \cite{efremov,ar} and from gluon topology 
\cite{bass99}, 
viz.
\begin{equation}
g_A^{(0)} 
= 
\Biggl(
\sum_q \Delta q - 3 {\alpha_s \over 2 \pi} \Delta g \Biggr)_{\rm partons}
+ \ {\cal C}
\end{equation}
Here ${1 \over 2} \Delta q$ and $\Delta g$ are the amount of spin 
carried by quark and gluon partons in the polarized proton and 
${\cal C}$ is the topological contribution.
The topological contribution is associated with Bjorken $x$ equal to
zero and is related to the role of zero-modes in axial U(1) symmetry
breaking.
It does not contribution to the value of $g_A^{(0)}|_{\rm pDIS}$ 
extracted from polarized deep inelastic scattering 
but does contribute to the value extracted from $\nu p$ elastic scattering.
A quality measurement of $\nu p$ elastic scattering could test theoretical
ideas about the axial U(1) problem and offers a real possibility to settle
the famous Crewther - 't Hooft debate \cite{rjc,thooft} through experiment.

A positive value of $\Delta g$ acts to reduce the value of 
$g_A^{(0)}|_{\rm pDIS}$  
extracted from polarized deep inelastic scattering.
Measuring the gluon polarization $\Delta g$ is one of the 
key goals of QCD spin physics \cite{gaby}.
A recent QCD motivated fit to the world $g_1$ data 
suggests a 
value $\Delta g = 0.63^{+0.20}_{-0.19}$ \cite{jechiel},
in agreement with the prediction \cite{bbs1} based on colour coherence and 
perturbative QCD.

The interplay between the proton spin problem and the U(1) problem is 
further manifest in the flavour-singlet Goldberger-Treiman relation 
\cite{venez} which connects $g_A^{(0)}$ with the $\eta'$--nucleon 
coupling constant $g_{\eta' NN}$.
Working in the chiral limit it reads
\begin{equation}
M g_A^{(0)} = \sqrt{3 \over 2} F_0 \biggl( g_{\eta' NN} - g_{QNN} \biggr) 
\end{equation}
where
$g_{\eta' NN}$ is the $\eta'$--nucleon coupling constant and $g_{QNN}$ 
is an OZI violating coupling which measures the one particle irreducible 
coupling of the topological charge density 
$Q = {\alpha_s \over 4 \pi} G {\tilde G}$ to the nucleon.
($M$ is the nucleon mass and $F_0$ ($\sim 0.1$GeV) 
 renormalises the flavour-singlet decay constant.)
It is important to look for other observables which are sensitive 
to $g_{QNN}$. Gluonic degrees of freedom induce a contact term in 
the low-energy $pN \rightarrow pN \eta'$ reaction with strength 
proportional to $g_{QNN}$ \cite{cracow,sb99}. 
The strength of this interaction is presently under experimental study at 
COSY \cite{cosyprop} .

In Section 2 we give a brief review of the QCD axial anomaly and its 
relation to $g_A^{(0)}$. 
In Section 3 we discuss QCD anomaly effects in 
the intrinsic and orbital contributions to the proton's spin.
Section 4 discusses $\nu p$ elastic scattering as a probe of U(1) 
symmetry breaking.
Finally, Section 5 highlights the relationship between the 
proton spin problem and possible OZI violation in 
$\eta'$--production in low-energy proton-proton collisions.

\section{GLUON TOPOLOGY AND $g_A^{(0)}$}

The flavour-singlet axial charge $g_A^{(0)}$ is measured by 
the proton forward matrix element of the gauge invariantly
renormalised axial-vector current
\begin{equation}
J^{GI}_{\mu5} = \biggl[ \bar{u}\gamma_\mu\gamma_5u
                  + \bar{d}\gamma_\mu\gamma_5d
                  + \bar{s}\gamma_\mu\gamma_5s \biggr]^{GI}_{\mu^2}
\end{equation} 
viz.
\begin{equation}
2m s_\mu g_A^{(0)} = 
\langle p,s|
\ J^{GI}_{\mu5} \ |p,s\rangle 
\end{equation}
In QCD the axial anomaly \cite{adler,bell} induces various gluonic
contributions to $g_A^{(0)}$.
The flavour-singlet axial-vector current satisfies the anomalous
divergence equation
\begin{equation}
\partial^\mu J^{GI}_{\mu5}
= 2f\partial^\mu K_\mu + \sum_{i=1}^{f} 2im_i \bar{q}_i\gamma_5 q_i
\end{equation}
where
\begin{equation}
K_{\mu} = {g^2 \over 16 \pi^2}
\epsilon_{\mu \nu \rho \sigma}
\biggl[ A^{\nu}_a \biggl( \partial^{\rho} A^{\sigma}_a 
- {1 \over 3} g 
f_{abc} A^{\rho}_b A^{\sigma}_c \biggr) \biggr]
\end{equation}
is a renormalised version of the Chern-Simons current
and $f=3$ is the number of light-flavours.
Eq.(6) allows us to write
\begin{equation}
J_{\mu 5}^{GI} = J_{\mu 5}^{\rm con} + 2f K_{\mu}
\end{equation}
where 
\begin{equation}
\partial^{\mu} K_{\mu} 
= {g^2 \over 32 \pi^2} G_{\mu \nu} {\tilde G}^{\mu \nu}
\end{equation}
and
\begin{equation}
\partial^\mu J^{\rm con}_{\mu5}
= \sum_{l=1}^{f} 2im_i \bar{q}_l\gamma_5 q_l
\end{equation}
The partially conserved axial-vector current $J_{\mu 5}^{\rm con}$ 
and 
the Chern-Simons current $K_{\mu}$ are separately gauge dependent.
When we make a gauge transformation $U$ the gluon field transforms 
as
\begin{equation}
A_{\mu} \rightarrow U A_{\mu} U^{-1} + {i \over g} (\partial_{\mu} U) U^{-1}
\end{equation}
and the operator $K_{\mu}$
transforms as
\begin{eqnarray}
K_{\mu} &\rightarrow& K_{\mu}  \\ \nonumber
&+& i {g \over 16 \pi^2} \epsilon_{\mu \nu \alpha \beta}
\partial^{\nu} 
\biggl( U^{\dagger} \partial^{\alpha} U A^{\beta} \biggr)
\\ \nonumber
&+& {1 \over 96 \pi^2} \epsilon_{\mu \nu \alpha \beta}
(U^{\dagger} \partial^{\nu} U) 
(U^{\dagger} \partial^{\alpha} U)
(U^{\dagger} \partial^{\beta} U) 
.
\end{eqnarray}
Gauge transformations shuffle a scale invariant operator 
quantity between the two operators $J_{\mu 5}^{\rm con}$ 
and $K_{\mu}$ whilst keeping $J_{\mu 5}^{GI}$ invariant.

To make contact with the QCD parton model one would like to isolate the 
gluonic contribution to $g_A^{(0)}$ associated with $K_{\mu}$ and thus 
write $g_A^{(0)}$ as the sum of ``quark'' and ``gluonic'' contributions.
Here we have to be careful because of the gauge dependence of $K_{\mu}$.

Whilst $K_{\mu}$ is a gauge dependent operator, its forward matrix elements 
are invariant under the ``small'' gauge transformations of perturbative QCD.
In deep inelastic processes the internal structure of the nucleon is
described by the QCD parton model.
The deep inelastic structure functions may be written as the sum over 
the convolution of ``soft'' quark and gluon parton distributions with 
``hard'' photon-parton scattering coefficients.
Working in light-cone gauge $A_+=0$ 
the (target dependent) parton distributions describe a flux of quark 
and gluon partons carrying some fraction
$x = p_{+ \rm parton} / p_{+ \rm proton}$
of the proton's momentum
into the hard (target independent) photon-parton interaction which is 
described by the hard scattering coefficients.
In the QCD parton model one finds \cite{efremov,ar,ccm,bint}
\begin{equation}
g_A^{(0)}|_{\rm partons} 
= 
\Biggl(
\sum_q \Delta q - 3 {\alpha_s \over 2 \pi} \Delta g \Biggr)_{\rm partons}
\end{equation}
Here, 
in perturbative QCD,
$\Delta q_{\rm partons}$ 
is measured by $J_{+5}^{\rm con}$ and
$\Delta g_{\rm partons}$ is measured by $K_+$ (in $A_+=0$ gauge).
The polarised gluon contribution to Eq.(13) is characterised by 
the contribution to the first moment of $g_1$ 
from two-quark-jet events carrying large transverse momentum squared
$k_T^2 \sim Q^2$ which are generated by photon-gluon fusion \cite{ccm,bbs1}.
The polarised quark contribution $\Delta q_{\rm parton}$ 
is associated with the first moment of the measured $g_1$ 
after these two-quark-jet events are subtracted from the total data set.
\footnote{
The parton model decomposition of $g_A^{(0)}|_{\rm partons}$ 
written in (2) and (13) is also obtained if we make a cut-off 
on the invariant mass-squared of the ${\bar q} q$ 
pair produced in polarized photon-gluon fusion \cite{bbs98},
and in the more formal 
`AB' \cite{ab} and `JET' \cite{jet} schemes.
In the `${\overline {\rm MS}}$' \cite{msbar}
factorization scheme $\Delta q_{\rm {\overline{MS}}} 
\equiv 
( \Delta q  - {\alpha_s \over 2 \pi} \Delta g )_{\rm partons, \ AB, \ JET}$.
}

The QCD parton model formula (13) is not the whole story. 
Choose a covariant gauge. 
When we go beyond perturbation theory, the forward matrix 
elements of $K_{\mu}$ are not invariant under ``large'' 
gauge transformations 
which change the topological winding number \cite{jaffem}.
The issue of ``large'' gauge transformations means that
the spin structure of the proton is especially sensitive 
to the details of axial U(1) symmetry breaking.

Spontaneous U(1) symmetry breaking in QCD is associated with a massless 
Kogut-Susskind pole which couples equally and with opposite sign to the 
two gauge dependent currents $J_{+5}^{\rm con}$ and $K_{\mu}$, 
thus decoupling from $J_{\mu 5}^{GI}$ \cite{rjc}.
Large gauge transformations shuffle the residue of 
this massless pole
between the $J_{+5}^{\rm con}$ and $K_{\mu}$ contributions to $g_A^{(0)}$.
To see this consider the nucleon matrix element of $J_{\mu 5}^{GI}$
\begin{equation}
\langle p,s|J^{GI}_{5 \mu}|p',s'\rangle 
= 2m \biggl[ {\tilde s}_\mu G_A (l^2) + l_\mu l.{\tilde s} G_P (l^2) \biggr]
\end{equation}
where $l_{\mu} = (p'-p)_{\mu}$
and
${\tilde s}_{\mu} 
= {\overline u}_{(p,s)} \gamma_{\mu} \gamma_5 u_{(p',s')} / 2m $.
Since $J^{GI}_{5 \mu}$ does not couple to a massless 
Goldstone boson 
(the $\eta'$ is heavy)
it follows that $G_A(l^2)$ and $G_P(l^2)$ contain
no massless pole terms.
The forward matrix element of $J^{GI}_{5 \mu}$ is well
defined and
$g_A^{(0)} = G_A (0)$.
In covariant gauge we can write
\begin{equation}
\langle p,s|K_\mu |p',s'\rangle 
= 2m \biggl[ {\tilde s}_\mu K_A(l^2) + l_\mu l.{\tilde s} K_P(l^2) \biggr]
\end{equation}
where $K_P$ contains the massless Kogut-Susskind pole.
This massless pole cancels with a corresponding massless
pole term in $(G_P - K_P)$.
We may define a gauge-invariant form-factor $\chi^{g}(l^2)$
for the topological charge density (9) in the divergence of 
$K_{\mu}$:
\begin{equation}
2m l.{\tilde s} \chi^g(l^2) =
\langle p,s | {g^2 \over 8 \pi^2} G_{\mu \nu} {\tilde G}^{\mu \nu}
 | p', s' \rangle.
\end{equation}
Working in a covariant gauge, we find
\begin{equation}
\chi^{g}(l^2) = K_A(l^2) + l^2 K_P(l^2)
\end{equation}
by contracting Eq.(15) with $l^{\mu}$.
When we make a gauge transformation any change 
$\delta_{\rm gt}$
in $K_A(0)$ is compensated
by a corresponding change in the residue of the Kogut-Susskind
pole in $K_P$, viz.
\begin{equation}
\delta_{\rm gt} [ K_A(0) ]
+ \lim_{l^2 \rightarrow 0} \delta_{\rm gt} [ l^2 K_P(l^2) ] = 0.
\end{equation}

Topological winding number is a non-local property of QCD.
It is determined by the gluonic boundary conditions at 
``infinity'' \cite{rjc}
---
 a large surface with boundary which is spacelike with respect 
 to the positions $z_k$ of any operators or fields in the physical
 problem ---
and is insensitive to any local deformations of the gluon field 
$A_{\mu}(z)$ or of the gauge transformation $U(z)$
--- that is, perturbative QCD degrees of freedom.
When we take the Fourier transform to momentum space the topological 
structure induces a light-cone zero-mode which has support 
only at $x=0$. 
Hence, we are led to consider the possibility that there may 
be a term in $g_1$ which is proportional to $\delta(x)$ \cite{bass98}.

It remains an open question whether the net non-perturbative quantity 
which is shuffled between the $J_{\mu 5}^{\rm con}$ and $K_{\mu}$ 
contributions to $g_A^{(0)}$ under ``large'' gauge transformations is 
finite or not.
If it is finite and, therefore, physical then we find a net 
topological contribution ${\cal C}$ to $g_A^{(0)}$ \cite{bass99}
\begin{equation}
g_A^{(0)} 
= 
\Biggl(
\sum_q \Delta q - 3 {\alpha_s \over 2 \pi} \Delta g \Biggr)_{\rm partons}
+ \ {\cal C}
\end{equation}
The topological term ${\cal C}$ has support only at $x=0$.
\footnote{
Possible $\delta(x)$ terms in deep inelastic structure functions 
are also found in Regge theory where they are induced by 
Regge fixed poles with non-polynomial residue \cite{fixedpoles}.
}
Topological $x=0$ polarization is inacessible to polarized deep inelastic 
scattering experiments which measure 
$g_1(x,Q^2)$ between some small but finite value 
$x_{\rm min}$ and an upper value $x_{\rm max}$ 
which is close to one.
As we decrease $x_{\rm min} \rightarrow 0$ we measure the first moment
\begin{equation}
\Gamma \equiv \lim_{x_{\rm min} \rightarrow 0} \ 
\int^1_{x_{\rm min}} dx \ g_1 (x,Q^2).
\end{equation}
This means that the singlet axial charge which is extracted
from polarized deep inelastic scattering is 
the combination $g_A^{(0)}|_{\rm pDIS} = (g_A^{(0)} - {\cal C})$.
In contrast, elastic ${\rm Z}^0$ exchange 
processes such as 
$\nu p$ elastic scattering 
measure the full $g_A^{(0)}$.
One can, in principle, measure the topology term ${\cal C}$ 
by comparing the flavour-singlet axial charges which are extracted 
from polarized deep inelastic and $\nu p$ elastic scattering experiments.
A decisive measurement of $\nu p$ elastic scattering 
may be possible with the MiniBooNE set-up at FNAL \cite{tayloe}.

\section{SPIN AND ANGULAR MOMENTUM IN THE TRANSITION 
         BETWEEN CURRENT TO CONSTITUENT QUARKS}

One of the most challenging problems in particle physics is to understand 
the transition between the fundamental QCD ``current'' quarks and gluons
and the constituent quarks of low-energy QCD.
Relativistic constituent-quark pion coupling models predict
$g_A^{(0)} \simeq 0.6$
(the OZI value and twice the value of $g_A^{(0)}|_{\rm pDIS}$ in Eq.(1)).
If some fraction of the spin of the constituent quark is 
carried by gluon topology in QCD,
then the constituent quark model predictions for $g_A^{(0)}$ 
are not necessarily in contradiction with the small value of 
$g_A^{(0)}|_{\rm pDIS}$ extracted from deep inelastic scattering experiments.

The quark total angular momentum $J_q$ measured through the proton 
matrix of the angular momentum tensor in QCD can, in principle, be 
extracted from deeply virtual Compton scattering \cite{jidvcs}.
This $J_q$ is anomaly free in 
both perturbative and non-perturbative QCD \cite{bassdvcs,shorew}.
This means that the axial anomaly cancels between the intrinsic and 
orbital contributions to $J_q$.
Furthermore, any zero-mode contributions 
to the 
``quark intrinsic spin'' 
associated with $g_A^{(0)}$
are cancelled by 
zero-mode contributions to the 
``quark orbital angular momentum'',
which is measured by the proton matrix element of 
$[{\bar q} ({\vec z} \ {\rm x} \ {\vec D})_3 q](0)$.
Future measurements of 
``quark orbital angular momentum'' 
from DVCS should be quoted with respect 
to the factorization scheme and
process 
(polarized deep inelastic scattering or $\nu p$
 elastic scattering)
used to extract information about the ``intrinsic spin''.

\section{INSTANTONS AND $U_A(1)$ SYMMETRY BREAKING}

The presence or absence of topological $x=0$ polarization is 
intimately 
related to the dynamics of $U_A(1)$ symmetry breaking in QCD.

A key issue in dynamical $U_A(1)$ symmetry breaking is the role of 
instantons.
Whether instantons spontaneously \cite{rjc} or explicitly 
\cite{thooft} 
break $U_A(1)$ symmetry depends on the role of zero-modes in 
the 
quark-instanton 
interaction and how one should include 
non-local structure into the local anomalous Ward identity, 
Eq. (6).
Both scenarios start from 't Hooft's observation 
\cite{thooftinst} 
that the flavour determinant
\begin{equation}
\langle {\rm det} \biggl[
{\rm {\overline q}_L}^i {\rm q_R}^j {\rm (z)} 
\biggr]
\rangle_{\rm inst.} \neq 0
\end{equation}
in the presence of a vacuum tunneling process between states 
with different topological winding number \cite{marga}.
(We denote the tunneling process by the subscript ``inst.''.
 It is not specified at this stage whether ``inst.'' denotes 
 an instanton or an anti-instanton.)
To go further one has to be precise how one defines chirality
in (21):
either through
$J_{\mu 5}^{GI}$ \cite{thooft}
or through $J_{\mu 5}^{\rm con}$ \cite{rjc}.

As we now explain the two choices lead to different phenomenology.

Quark instanton interactions flip quark ``chirality'' 
so that 
when we time average over multiple scattering on an ensemble 
of instantons and anti-instantons the spin asymmetry measured
in polarized deep inelastic scattering is reduced relative to 
the asymmetry one would measure if instantons were not important.
Topological $x=0$ polarization is natural \cite{bass99,bass98} 
in the spontaneous symmetry breaking scenario where any instanton 
induced suppression of $g_A^{(0)}|_{\rm pDIS}$ is compensated 
by a shift of flavour-singlet axial-charge from partons carrying 
finite momentum $x > 0$ to a zero-mode at $x=0$ so that the total 
$g_A^{(0)}$ is conserved.
It is not generated by the explicit symmetry breaking scenario 
where 
the total 
$g_A^{(0)}$ 
(rather than the chirality measured by $J_{\mu 5}^{\rm con}$)
is sensitive to the quark-instanton interaction.
Comparing the values of $g_A^{(0)}$ extracted 
from $\nu p$ elastic and polarized deep inelastic scattering 
could provide valuable information on $U_A(1)$ symmetry breaking in QCD.

\begin{figure}[htb]
\includegraphics[width=\columnwidth]{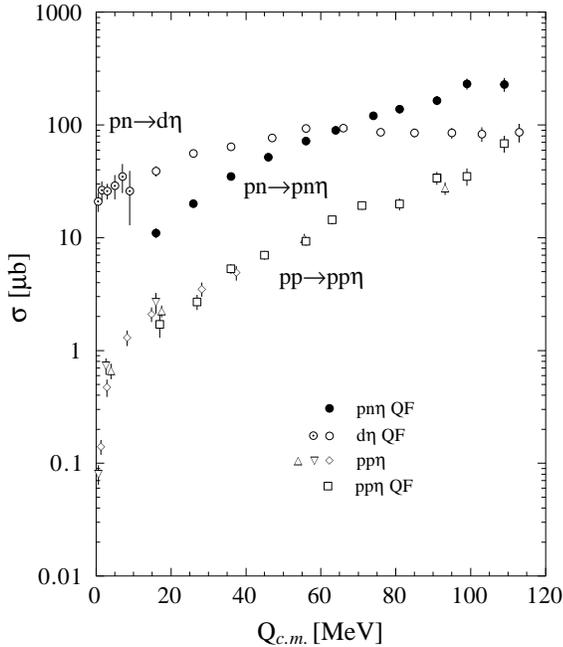}
\caption{The CELSIUS data on 
$pp \rightarrow pp \eta$ and 
$pn \rightarrow pn \eta$.}
\label{fig:lcsr}
\end{figure}

\section{PROTON SPIN STRUCTURE AND THE $\eta'$-NUCLEON INTERACTION}

Motivated by the flavour-singlet Goldberger-Treiman relation (3),
further insight into 
the relationship between the proton's internal 
spin structure and $U_A(1)$ dynamics may come 
from studying possible OZI violation in the $\eta'$--nucleon system.

Working with the $U_A(1)$--extended chiral Lagrangian for low-energy 
QCD \cite{vecca} one finds a gluon-induced contact interaction in 
the $pp \rightarrow pp \eta'$ reaction close to threshold \cite{sb99}:
\begin{equation}
{\cal L}_{\rm contact} =
         - {i \over F_0^2} \ g_{QNN} \ {\tilde m}_{\eta_0}^2 \
           {\cal C} \
           \eta_0 \ 
           \biggl( {\bar p} \gamma_5 p \biggr)  \  \biggl( {\bar p} p \biggr)
\end{equation}
Here ${\tilde m}_{\eta_0}$ is the gluonic contribution to the mass of the
singlet 0$^-$ boson and ${\cal C}$ is a second OZI violating coupling 
which also features in $\eta'N$ scattering.
The physical interpretation of the contact term (22) 
is a ``short distance'' ($\sim 0.2$fm) interaction 
where glue is excited in the interaction region of
the proton-proton collision and 
then evolves to become an $\eta'$ in the final state.
This gluonic contribution to the cross-section 
for $pp \rightarrow pp \eta'$ 
is extra to the contributions associated with meson exchange models
\cite{faldt}. 
There is no reason, a priori, to expect it to be small.

Since glue is flavour-blind the contact interaction (22) has the same 
size in both 
the $pp \rightarrow pp \eta'$ and $pn \rightarrow pn \eta'$ reactions.
CELSIUS \cite{celsius} have measured the ratio
$R_{\eta} 
 = \sigma (pn \rightarrow pn \eta ) / \sigma (pp \rightarrow pp \eta )$
for quasifree $\eta$ 
production from a deuteron target up to 100 MeV above threshold.
They observed that $R_{\eta}$ is approximately energy-independent 
$\simeq 6.5$ over the whole energy range --- see Fig.1.
The value of this ratio signifies a strong isovector exchange 
contribution to the $\eta$ production mechanism \cite{celsius}.
This experiment should be repeated for $\eta'$ production.
The cross-section for $pp \rightarrow pp \eta'$ 
close to threshold has been measured at COSY \cite{cosy}.
Following the suggestion in \cite{sb99} new experiments 
\cite{cosyprop} at COSY have been initiated to carry out 
the $pn \rightarrow pn \eta'$ measurement.
The more important that the gluon-induced process (22) is in 
the $pp \rightarrow pp \eta'$ reaction the more one would expect 
$R_{\eta'} =
 \sigma (pn \rightarrow pn \eta' ) / \sigma (pp \rightarrow pp \eta' )$
to approach unity near threshold
after we correct for the final state 
interaction \cite{faldt} between the two outgoing nucleons.
(After we turn on the quark masses, 
 the small $\eta-\eta'$ mixing angle 
 $\theta \simeq -18$ degrees means that the gluonic effect
 (22) should be considerably bigger in $\eta'$ production than $\eta$
 production.) 
$\eta'$ phenomenology is characterised by large OZI violations.
It is natural to expect 
large gluonic effects in the $pp \rightarrow pp \eta'$ process.

\vspace{3ex}

I thank R. L. Jaffe and A. W. Thomas for helpful conversations.

\end{document}